\documentclass[preprint,prd,aps,showpacs,showkeys,nofootinbib]{revtex4}
\usepackage{graphicx}
\usepackage{dcolumn}
\usepackage{bm}
\begin{document}

\title{ Electric
dipole moments of neutron and heavy quarks c, t in CP violating  $U(1)_X$SSM}

\author{Ben Yan$^{1,2*}$\footnote{email:yb118sdfz@163.com}, Shu-Min Zhao$^{1,2*}$\footnote{email:zhaosm@hbu.edu.cn}, Tai-Fu Feng$^{1,2,3}$\footnote{email:fengtf@hbu.edu.cn}
\footnotetext{* Indicates equal contribution}
}

\affiliation{$^1$ Department of Physics, Hebei University, Baoding 071002, China}
\affiliation{$^2$ Key Laboratory of High-precision Computation and Application of Quantum Field Theory of Hebei Province, Baoding 071002, China}
\affiliation{$^3$ Department of Physics, Chongqing University, Chongqing 401331, China}
\date{\today}

\begin{abstract}
We study and analyze the electric dipole moments (EDMs) of neutron and heavy quarks c, t in a CP-violating (CPV) supersymmetric U(1) extension of the standard model called $U(1)_X$SSM, whose local gauge group is $U(1)_X\times SU(3)_C\times SU(2)_L \times U(1)_Y$.
The contributions from the one loop diagrams and the Weinberg operators are taken into account with the new introduced CPV phases beyond the minimum supersymmetric standard model(MSSM) being non-zero.
Our numerical results show that in $U(1)_X$SSM the neutron EDM can be smaller than its experimental upper bound $1.8\times10^{-26}$ e$\cdot$cm without a small CPV phase.
The magnitudes of EDMs of heavy quarks c and t can reach $10^{-24}$ e$\cdot$cm. and $10^{-22}$ e$\cdot$cm respectively. This is favorable to the study of CPV.
\end{abstract}

\pacs{\emph{}}

\keywords{CP violating, neutron, electric dipole moment}

\maketitle

\section{Introduction}
One of the outstanding problem of particle physics today is the origin of CPV. The CPV found in the K and B system \cite{In0,In1,In2,In2plus} can be well explained in the standard model(SM). In order to research the CPV sources, it is important to find other processes which also violate the CP conservation. The EDMs of elementary particles, especially neutrons, are clear signal of CPV\cite{In3plus,In3,In4}.

In the SM, the phase in Cabbibo-Kobayashi-Maskawa(CKM) matrix is the only source of CPV because of  one scalar doublet\cite{In5,In5plus}. The EDM of a fermion not only disappear even up to two loop order, but also there are partial cancelation between the three loop contributions\cite{In6,In7,In8}. It means that, if a large EDM of an elementary fermion is detected, one can confirm that there exists new physics(NP) beyond the SM. The measurements of the neutron EDM($d_n$) possesses a very stringent upperbound. The latest experimental results show that $|d_n|<1.8\times10^{-26}$ e$\cdot$cm (90\% C.L.) \cite{In9}, while  the SM predicts a much smaller $|d_n|<10^{-31}$ e$\cdot$cm\cite{smedm1,smedm2}. There are similar bounds on the c quark and t quark EDMs $|d_c|<5.0\times10^{-17}$ e$\cdot$cm and $|d_t|<2.17\times10^{-16}$ e$\cdot$cm\cite{smedm3plus,smedm3}. Furthermore,  the indirect constraints on c quark and t quark EDMs can be improved to $|d_c|<1.5\times10^{-21}$ e$\cdot$cm and $|d_t|<5.0\times10^{-20}$ e$\cdot$cm\cite{edmdc,edmdt}. These upper bounds of heavy quarks are much larger than that of neutron, and they may be detected in the near future.

Although the SM has achieved incredible success, it is unable to explain many NP phenomena. Physicists consider that the SM should be a low energy and effective theory of a large model. Supersymmetric(SUSY) theories are perhaps the most widely considered extensions to the SM. SUSY models remove fine-tunings present in quadratically divergent radiative corrections to the Higgs mass and introduce new CPV sources from the complex phases of the new parameters\cite{susy1,susy1.1,susy1.2,susy1.3,susy1.4}. In MSSM, the new CPV sources which contribute to the EDMs of quarks and leptons were widely studied by people. Unfortunately, for moderate mass scales, MSSM predicts a large neutron EDM $d_n$ of order $(10^{-22}\sim 10^{-23})\tilde{\phi}$ e$\cdot$cm\cite{susy2}, where $\tilde{\phi}$ is some combination of SUSY phases. There are two approaches to satisfy the experimental limit if the CP phases are not sufficient small\cite{susy3,susy3plus}. One is the fine-tuning: various contributions cancel with each other in some special parameter spaces, and the other is to make the SUSY particles very heavy at several TeV order. There are also other works of the EDMs for neutron and heavy quarks\cite{OCEDM1,OCEDM2,OCEDM3}.

The $U(1)_X$SSM is a popular $U(1)$ extension of MSSM. In $U(1)_X$SSM, the new superfields beyond MSSM are three $SU(2)_L$ singlet Higgs superfields $\hat{\eta},~\hat{\bar{\eta}},~\hat{S}$, three right-handed neutrinos and a gauge boson.
The $U(1)_X$SSM has several advantages than the MSSM: light neutrinos can obtain tiny mass; the lightest scalar neutrino possesses cold dark matter characters; the $\mu$ problem and the little hierarchy problem in MSSM are both relieved, because $U(1)_X$SSM has the $\lambda_H\hat{S}\hat{H}_u\hat{H}_d$ term and more superfields; the mass of the lightest CP-even Higgs at the tree level is enhanced by the addition of the three singlet Higgs superfields, and the correction of the stop to Higgs mass does not need to be very large.

In $U(1)_X$SSM, there are several new CPV sources beyond the MSSM: 1 the gaugino mass of $U(1)_X$; 2 the mixing mass of $U(1)_X$ gaugino and $U(1)_Y$ gaugino;
 3 the mass of the new higgsino. Researching the effects of the new CPV sources on neutron EDM and heavy quarks(c, t) EDM can open a new window for the study of the CPV and NP beyond SM. 
 Our work shows that in $U(1)_X$SSM, the neutron EDM affected by the new introduced CPV phase can satisfy the strong constraint of $d_n$ without requiring a very small phase angle value as in MSSM.

The paper is organized as follows.
In Section II, the detailed introduction of $U(1)_X$SSM can be found.
In section III, we research the EDMs of heavy c,t quarks and neutron from the one loop diagrams and the Weinberg operator contributions.
In section IV, we show the numerical results and the derivation of reasonable parameter space. Our conclusion is summarized in section V.

\section{The $U(1)_X$SSM}
$U(1)_X$SSM is a non-universal Abelian extension of the MSSM and its local gauge group is extended to $SU(3)_C\times SU(2)_L \times U(1)_Y \times U(1)_X$. The addition of new $U(1)_X$ gauge symmetry implies that one needs to cancel the axial-vector anomaly and the mixed gravitational-gauge anomaly \cite{anomalyfree1,anomalyfree1plus,anomalyfree2}.
The $U(1)_X$SSM has been proven to be anomaly free. The detail of the anomaly cancellation in $U(1)_X$SSM can be found in Ref.\cite{anomalyfree3}. To construct this model, three singlet new Higgs superfields  $\hat{\eta},~\hat{\bar{\eta}},~\hat{S}$ and right-handed neutrinos $\hat{\nu}_i$ are added to MSSM. At the tree level, it can give light neutrino tiny mass  through the seesaw mechanism.
The particle content and charge assignments for $U(1)_X$SSM are shown in the table \ref{quarks}.
\begin{table}
	\caption{ The superfields in $U(1)_X$SSM}
	\begin{tabular}{|c|c|c|c|c|c|c|c|c|c|c|c|c|c|c|}
		\hline
		Superfields &~~$\hat{Q}_i$~~ & ~~$\hat{u}^c_i$~~ &~~$\hat{d}^c_i$~~ &~~$\hat{L}_i$~~&~~$\hat{e}^c_i$&~~$\hat{\nu}_i$~ ~&~~$\hat{H}_u$~~&~~$\hat{H}_d$~~&~~$\hat{\eta}$~~&~~$\hat{\bar{\eta}}$~~&~~$\hat{S}$~~ \\
		\hline
		$SU(3)_C$ & 3 & $\bar{3}$ & $\bar{3}$ & 1&1&1&1&1&1&1&1 \\
		\hline
		 $SU(2)_L$ & 2 & 1 & 1 & 2&1&1&2&2&1&1&1 \\
		\hline
		$U(1)_Y$ & 1/6 & -2/3 & 1/3 & -1/2&1&0&1/2&-1/2&0&0&0 \\
		\hline
		$U(1)_X$ & 0 & -1/2 &1/2 & 0 &1/2&-1/2&1/2&-1/2&-1&1&0 \\
		\hline
	\end{tabular}
	\label{quarks}
\end{table}
The superpotential of $U(1)_X$SSM is given by:
\begin{eqnarray}
&&W=l_W\hat{S}+\mu\hat{H}_u\hat{H}_d+M_S\hat{S}\hat{S}-Y_d\hat{d}\hat{q}\hat{H}_d-Y_e\hat{e}\hat{l}\hat{H}_d+\lambda_H\hat{S}\hat{H}_u\hat{H}_d
\nonumber\\&&+\lambda_C\hat{S}\hat{\eta}\hat{\bar{\eta}}+\frac{\kappa}{3}\hat{S}\hat{S}\hat{S}+Y_u\hat{u}\hat{q}\hat{H}_u+Y_X\hat{\nu}\hat{\bar{\eta}}\hat{\nu}
+Y_\nu\hat{\nu}\hat{l}\hat{H}_u.
\end{eqnarray}
The soft SUSY breaking terms are
\begin{eqnarray}
&&\mathcal{L}_{soft}=\mathcal{L}_{soft}^{MSSM}-B_SS^2-L_SS-\frac{T_\kappa}{3}S^3-T_{\lambda_C}S\eta\bar{\eta}
+\epsilon_{ij}T_{\lambda_H}SH_d^iH_u^j\nonumber\\&&
-T_X^{IJ}\bar{\eta}\tilde{\nu}_R^{*I}\tilde{\nu}_R^{*J}
+\epsilon_{ij}T^{IJ}_{\nu}H_u^i\tilde{\nu}_R^{I*}\tilde{l}_j^J
-m_{\eta}^2|\eta|^2-m_{\bar{\eta}}^2|\bar{\eta}|^2-m_S^2S^2\nonumber\\&&
-(m_{\tilde{\nu}_R}^2)^{IJ}\tilde{\nu}_R^{I*}\tilde{\nu}_R^{J}
-\frac{1}{2}\Big(M_{BL}\lambda^2_{\tilde{X}}+2M_{BB^\prime}\lambda_{\tilde{B}}\lambda_{\tilde{X}}\Big)+h.c.~~.
\end{eqnarray}

The two Abelian groups $U(1)_Y$ and $U(1)_X$ in $U(1)_X$SSM
have a new effect called as the gauge kinetic mixing comparing with just one Abelian gauge group model such as the MSSM.
As the Abelian gauge group  $U(1)_Y$ extends to two Abelian groups $ U(1) _Y\otimes U(1) _{X} $, the covariant derivatives correspondingly take a matrix form of  $D_{\mu}=\partial_{\mu}-iQ_{\phi }^TGA'$. For the two Abelian gauge groups are unbroken, one can perform a basis transformation to absorb the $U(1)_Y$ and  $U(1)_X$ gauge kinetic mixing. It is the covariant derivative redefinition: $Q_{\phi }^TG(R^T R)A'=Q_{\phi }^T \tilde{ G}A$. The matrices $G$, $\tilde{G}$, $A$, $A'$ are
\vspace{0.3cm}
$G=\left(\begin{array}{cc}g_{Y} & g{'}_{YX}\\g{'}_{XY} &g{'}_{X}\end{array}\right)$, $\tilde{G}=\left(\begin{array}{cc}g_1 & g_{YX} \\0 & g_X\end{array}\right)$,
$A=\left(\begin{array}{c}A_{\mu}^{Y} \\ A_{\mu}^{X}\end{array}\right)$ and
$A'=\left(\begin{array}{c}A_{\mu}^{\prime Y} \\ A_{\mu}^{\prime X}\end{array}\right)
$ respectively, where $A_{\mu}^{\prime Y}$ and $A^{\prime X}_\mu$ denote the gauge fields of $U(1)_Y$ and $U(1)_X$,
and $Y$ and $X$ represent the hypercharge and $X$ charge respectively.

The gauge bosons $A^{X}_\mu,~A^Y_\mu$ and $V^3_\mu$ mix together at the tree level. In the basis $(A^Y_\mu, V^3_\mu, A^{X}_\mu)$, their mass matrix
is written as\cite{g-2}
\begin{eqnarray}
&&\left(\begin{array}{*{20}{c}}
\frac{1}{8}g_{1}^2 v^2 &~~~ -\frac{1}{8}g_{1}g_{2} v^2 & ~~~\frac{1}{8}g_{1}(g_{{YX}}+g_X) v^2 \\
-\frac{1}{8}g_{1}g_{2} v^2 &~~~ \frac{1}{8}g_{2}^2 v^2 & ~~~~-\frac{1}{8}g_{2}(g_{{YX}}+g_X) v^2\\
\frac{1}{8}g_{1}(g_{{YX}}+g_X) v^2 &~~~ -\frac{1}{8}g_{2}(g_{{YX}}+g_X) v^2 &~~~~ \frac{1}{8}(g_{{YX}}+g_X)^2 v^2+\frac{1}{8}g_{{X}}^2 \xi^2
\end{array}\right),\label{gauge matrix}
\end{eqnarray}
with $v^2=v_u^2+v_d^2$ and $\xi^2=v_\eta^2+v_{\bar{\eta}}^2$. It is convenient to use an unitary matrix including two mixing angles $\theta_{W}$ and $\theta_{W}'$ to diagonalize this mass matrix in Eq. (\ref{gauge matrix}),
\begin{eqnarray}
&&\left(\begin{array}{*{20}{c}}
\gamma_\mu\\ [6pt]
Z_\mu\\ [6pt]
Z'_\mu
\end{array}\right)=
\left(\begin{array}{*{20}{c}}
\cos\theta_{W} & \sin\theta_{W} & 0 \\ [6pt]
-\sin\theta_{W}\cos\theta_{W}' & \cos\theta_{W}\cos\theta_{W}' & \sin\theta_{W}'\\ [6pt]
\sin\theta_{W}\sin\theta_{W}' & -\cos\theta_{W}'\sin\theta_{W}' & \cos\theta_{W}'
\end{array}\right)
\left(\begin{array}{*{20}{c}}
A^Y_\mu\\ [6pt]
V^3_\mu\\ [6pt]
A^{X}_\mu
\end{array}\right).
\end{eqnarray}
$\theta_{W}$ is the Weinberg angle and $\theta_{W}^\prime$ is the new mixing angle appearing in the couplings of $Z$ and $Z^{\prime}$, which is
defined as\cite{g-2}
\begin{eqnarray}
\sin^2\theta_{W}'=\frac{1}{2}-\frac{[(g_{{YX}}+g_X)^2-g_{1}^2-g_{2}^2]v^2+
	4g_{X}^2\xi^2}{2\sqrt{[(g_{{YX}}+g_X)^2+g_{1}^2+g_{2}^2]^2v^4+8g_{X}^2[(g_{{YX}}+g_X)^2-g_{1}^2-g_{2}^2]v^2\xi^2+16g_{X}^4\xi^4}}.
\end{eqnarray}

Then, three particles $M_Z$, $M_{Z'}$ and a massless photon can be obtained from the diagonalized gauge boson matrix. The square masses of $M_Z$ and $M_{Z'}$ are,
\begin{eqnarray}\nonumber
&&M_{Z,Z'}^2 = \frac{1}{8}\Big( {[g_1^2 + g_2^2 + (g_{{YX}}+g_X)^2]{v^2} + 4g_X^2{\xi ^2}}\nonumber \\
&&{\rm} { \mp \sqrt {{{[g_1^2 + g_2^2 + (g_{{YX}}+g_X)^2]}^2}{v^4} + 8[(g_{{YX}}+g_X)^2 - g_1^2 - g_2^2]g_X^2{v^2}{\xi ^2} + 16g_X^4{\xi ^4}} } \Big).
\end{eqnarray}

In the Higgs sector, there are two Higgs doublets and three Higgs singlets. Their explicit forms are shown in the follow,
\begin{eqnarray}
&&H_{u}=\left(\begin{array}{c}H_{u}^+\\{1\over\sqrt{2}}\Big(v_{u}+H_{u}^0+iP_{u}^0\Big)\end{array}\right),
~~~~~~
H_{d}=\left(\begin{array}{c}{1\over\sqrt{2}}\Big(v_{d}+H_{d}^0+iP_{d}^0\Big)\\H_{d}^-\end{array}\right),
\nonumber\\
&&\eta={1\over\sqrt{2}}\Big(v_{\eta}+\phi_{\eta}^0+iP_{\eta}^0\Big),~~~~~~~~~~~~~~~
\bar{\eta}={1\over\sqrt{2}}\Big(v_{\bar{\eta}}+\phi_{\bar{\eta}}^0+iP_{\bar{\eta}}^0\Big),\nonumber\\&&
\hspace{4.0cm}S={1\over\sqrt{2}}\Big(v_{S}+\phi_{S}^0+iP_{S}^0\Big).
\end{eqnarray}
$v_u,~v_d,~v_\eta$,~ $v_{\bar\eta}$ and $v_S$ are the corresponding  vacuum expectation values(VEVs) of the Higgs superfields $H_u$, $H_d$, $\eta$, $\bar{\eta}$ and $S$. Here, we define $\tan\beta=v_u/v_d$ and $\tan\beta'=v_{\bar{\eta}}/v_{\eta}$. The neutral CP-even parts of $H_u,~ H_d,~\eta,~\bar{\eta}$,~$S$ mix together and form a $5\times5 $ mass squared matrix. Note that the loop corrections to the lightest CP-even 
Higgs mass are important\cite{LCTHiggs1,LCTHiggs2}. In this research, they are taken into account in numerical calculation to coincide the Higgs mass of 125 GeV. Further detail relating to the $U(1)_X$SSM can be found in the literature \cite{anomalyfree3,U1X1} or use SARAH to generate a complete one \cite{U1X2}.
\section{The EDMs of neutron and heavy quarks}
\subsection{Mass matrices and couplings used to calculate EDM in $U(1)_X$SSM}
In $U(1)_X$SSM, the mass matrix of neutralino contains new CPV sources beyond the MSSM.
 Therefore, we specify the matrix here.
 In  the base $(\lambda_{\tilde{B}}, \tilde{W}^0, \tilde{H}_d^0, \tilde{H}_u^0,
\lambda_{\tilde{X}}, \tilde{\eta}, \tilde{\bar{\eta}}, \tilde{s}) $,
\begin{equation}
M_{\tilde{\chi}^0} = \left(
\begin{array}{cccccccc}
M_1 &0 &-\frac{g_1}{2}v_d &\frac{g_1}{2}v_u &{M}_{B B'} &0  &0  &0\\
0 &M_2 &\frac{1}{2} g_2 v_d  &-\frac{1}{2} g_2 v_u  &0 &0 &0 &0\\
-\frac{g_1}{2}v_d &\frac{1}{2} g_2 v_d  &0
&m_{\tilde{H}_d^0\tilde{H}_u^0} &m_{\tilde{H}_d^0\lambda_{\tilde{X}}} &0 &0 & - \frac{{\lambda}_{H} v_u}{\sqrt{2}}\\
\frac{g_1}{2}v_u &-\frac{1}{2} g_2 v_u  &m_{\tilde{H}_d^0\tilde{H}_u^0} &0 &m_{\tilde{H}_u^0\lambda_{\tilde{X}}} &0 &0 &- \frac{{\lambda}_{H} v_d}{\sqrt{2}}\\
{M}_{B B'} &0 &m_{\tilde{H}_d^0\lambda_{\tilde{X}}} &m_{\tilde{H}_u^0\lambda_{\tilde{X}}} &{M}_{BL} &- g_{X} v_{\eta}  &g_{X} v_{\bar{\eta}}  &0\\
0  &0 &0 &0 &- g_{X} v_{\eta}  &0 &\frac{1}{\sqrt{2}} {\lambda}_{C} v_S  &\frac{1}{\sqrt{2}} {\lambda}_{C} v_{\bar{\eta}} \\
0  &0 &0 &0 &g_{X} v_{\bar{\eta}}  &\frac{1}{\sqrt{2}} {\lambda}_{C} v_S  &0 &\frac{1}{\sqrt{2}} {\lambda}_{C} v_{\eta} \\
0 &0 & - \frac{{\lambda}_{H} v_u}{\sqrt{2}} &- \frac{{\lambda}_{H} v_d}{\sqrt{2}} &0 &\frac{1}{\sqrt{2}} {\lambda}_{C} v_{\bar{\eta}}
 &\frac{1}{\sqrt{2}} {\lambda}_{C} v_{\eta}  &m_{\tilde{s}\tilde{s}}\end{array}
\right),\label{mx0}
 \end{equation}
with
\begin{eqnarray}
&& m_{\tilde{H}_d^0\tilde{H}_u^0} = - \frac{1}{\sqrt{2}} {\lambda}_{H} v_S  - \mu ,~~~~~~~
m_{\tilde{H}_d^0\lambda_{\tilde{X}}} = -\frac{1}{2} (g_{Y X} + g_{X})v_d, \nonumber\\&&
m_{\tilde{H}_u^0\lambda_{\tilde{X}}} = \frac{1}{2} (g_{Y X} + g_{X})v_u
 ,~~~~~~~~~~~~
m_{\tilde{s}\tilde{s}} = 2 M_S  + \sqrt{2} \kappa v_S.\label{neutralino1}
\end{eqnarray}
Moreover, the chargino mass matrix $M_{\tilde{\chi}^\pm}$,
down type scalar quark squared mass matrix $M^2_{\tilde{D}}$ and  up type scalar quark squared mass matrix
 $M^2_{\tilde{U}}$ are needed in calculating the EDMs and CEDMs.
 They are similar to the those in MSSM.  We show them in the appendix.

The parameters ${M}_{1}$, ${M}_{2}$, $\mu$, ${M}_{BL}$, ${M}_{BB'}$ and ${M}_{S}$ can be complex.
Their phase angles can lead
	to the emerging of the CPV effect such as the EDM of fermion. Comparing with MSSM,
	$U(1)_X$SSM has the new parameters ${M}_{BB'}$, ${M}_{BL}$ and  ${M}_{S}$.

In $U(1)_X$SSM, the neutralino-quark-squark couplings  are different from the corresponding terms in MSSM.
We show their concrete forms here,
   \begin{eqnarray}
&&\mathcal{L}_{\chi^0d\tilde{D}}=-\frac{i}{6}\bar{\chi}^0_i\Big\{\Big[\sqrt{2}( g_1 Z_N^{1,i}
-3 g_2 Z_N^{2,i}   + g_{Y X} Z_N^{5,i}) (Z_{\tilde{D}}^*)^{{j,k}} +6 Z_N^{3,i} Y_d^j (Z_{\tilde{D}}^*)^{{3 + j,k}}   \Big]P_L\nonumber\\
&&+\Big[6 Y_d^j   (Z_{\tilde{D}}^*)^{{j,k}}  (Z_N^*)^{{3,i}}
+ \sqrt{2} (Z_{\tilde{D}}^*)^ {{3 + j,k}} [2 g_1 (Z_N^*)^{{1,i}}
 + (2 g_{Y X}  + 3 g_{X})(Z_N^*)^{{5,i}}]\Big]P_R\Big\}d_j\tilde{D}^*_k,\label{LX01}
\\
&&\mathcal{L}_{\chi^0u\tilde{U}}=-\frac{i}{6}\bar{\chi}^0_i\Big\{ \Big[\sqrt{2}( g_1 Z_N^{1,i}
+3  g_2 Z_N^{2,i}   + g_{Y X} Z_N^{5,i}   ) (Z_{\tilde{U}}^*)^{{j,k}}+6 Z_N^{4,i} Y_u^j (Z_{\tilde{U}}^*)^{{3 + j,k}}   \Big]P_L\nonumber\\
&&- \Big[ \sqrt{2}(Z_{\tilde{U}}^*)^{{3 + j,k}}  \Big((3 g_{X}  + 4 g_{Y X})(Z_N^*)^{{5,i}}
 + 4 g_1 (Z_N^*)^{{1,i}}\Big) -6 Y_u^j (Z_{\tilde{U}}^*)^{{j,k}} (Z_N^*)^{{4,i}}\Big] P_R\Big\}u_j\tilde{U}^*_k.\label{LX02}
\end{eqnarray}

Here $P_L=\frac{1-\gamma^5}{2}$, $P_R=\frac{1+\gamma^5}{2}$, $Z_N$,  $Z_{\tilde{D}}$ and $Z_{\tilde{U}}$ are the diagonalizing matrices for $M_{\tilde{\chi}^0}$,  $M^2_{\tilde{D}}$ and  $M^2_{\tilde{U}}$. They satisfy the relations $Z_N^{T} M_{\tilde{\chi}^0} Z_N=diag(m_{\tilde{\chi}^0_1},m_{\tilde{\chi}^0_2},...,m_{\tilde{\chi}^0_8})$
and  $Z_{\tilde{D}(\tilde{U})}^{\dagger} M^2_{\tilde{D}(\tilde{U})} Z_{\tilde{D}(\tilde{U})}=diag(m^2_{\tilde{D}(\tilde{U})_1},m^2_{\tilde{D}(\tilde{U})_2},...,m^2_{\tilde{D}(\tilde{U})_6})$, where
$m_{\tilde{\chi}^0_i}(i=1,2,...,8)$ and $m^2_{\tilde{D}(\tilde{U})_i}(i=1,2,...,6)$ denote the
corresponding mass eigenvalues of $M_{\tilde{\chi}^0}$ and $M^2_{\tilde{D}(\tilde{U})}$.

\subsection{Calculate the EDM}
The effective Lagrangian for the EDM $d_{f}$ of the fermion is defined through the dimension five operator\cite{EDM1},
\begin{eqnarray}
&&{\cal L}_{{EDM}}=-{i\over2}d^{\gamma}\overline{f}\sigma^{\mu\nu}\gamma_5
fF_{{\mu\nu}},
\label{eq1}
\end{eqnarray}
with $F_{{\mu\nu}}$ representing the electromagnetic field strength, $f$ denoting a fermion field.
Obviously, Eq.(\ref{eq1}) violates CP conservation, which can not be produced from the fundamental interactions at tree level.
The one loop diagrams may give non-zero contributions to $d_f$ in the CPV electroweak theory.
 Besides the operator in Eq.(\ref{eq1}), the chromoelectric dipole moment (CEDM) of quarks can also give contributions
  \begin{eqnarray}
&&{\cal L}_{{CEDM}}=-{i\over2}d^{g}\overline{q}T^a\sigma^{\mu\nu}\gamma_5
qG^a_{{\mu\nu}}.
\label{eq11}
\end{eqnarray}

We use the effective method to obtain the effective Lagrangian with the CPV operators at matching
scale $\Lambda$ which should be evolved down to the quark mass scale using the
renormalization group equations(RGEs). The effective Lagrangian with a full set of CPV operators relevant to the quark EDM and CEDM are,
\begin{eqnarray}
&&{\cal L}_{{eff}}=\sum\limits_{i}^5C_{i}(\Lambda){\cal O}_{i}(\Lambda)\;,
\nonumber\\
&&{\cal O}_{1}=\overline{q}\sigma^{\mu\nu}P_{L}qF_{{\mu\nu}}
\;,~~~~~{\cal O}_{2}=\overline{q}\sigma^{\mu\nu}P_{R}qF_{{\mu\nu}}
\;,~~~{\cal O}_{3}=\overline{q}T^a\sigma^{\mu\nu}P_{L}qG^a_{{\mu\nu}}
\;,\nonumber\\
&&{\cal O}_{4}=\overline{q}T^a\sigma^{\mu\nu}P_{R}qG^a_{{\mu\nu}}
\;,~{\cal O}_{5}=-{1\over6}f_{{abc}}G_{{\mu\rho}}^aG^{b\rho}_{\nu}
G_{{\lambda\sigma}}^c\epsilon^{\mu\nu\lambda\sigma}\;.
\label{eq3}
\end{eqnarray}
Here, $C_{i}(\Lambda)$ are the Wilson coefficients.

\begin{figure}[ht]
	\centering
		\includegraphics[width=100mm]{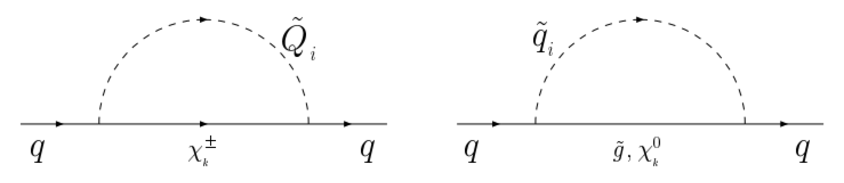}
		\caption{The one loop self-energy diagrams which lead to the quark EDMs and CEDMs.}
		\label{fig:Selfenergy}
\end{figure}

Thereafter we need to calculate the triangle diagrams for the quark EDMs and CEDMs. This can be obtained by attaching a photon(EDM) or gluon(CEDM) external line in all possible ways to the quark self-energy diagrams. The authors in Ref.\cite{EDM1} study the SUSY corrections to the quark EDMs and CEDMs.
In $U(1)_X$SSM, the one loop quark self-energy diagrams contributing to the quark EDMs and CEDMs are similar as those in MSSM. They are shown in Fig.\ref{fig:Selfenergy}.

Based on the Feynman diagrams,
the one loop contributions from the neutralino-squark diagrams to the quark EDMs and CEDMs are derived here. The results for up quarks are
\begin{eqnarray}
&&
d_{\chi_{k}^0}^\gamma(u^I)={e\over48\pi^2} \sum\limits_{i=1}^6\sum\limits_{k=1}^8{\bf Im}\Big(
(A_{L}^U)_{k,i}(A_R^U)_{i,k}^\dagger\Big)
{m_{\chi_{k}^ 0}\over m_{\tilde{U}_i}^2}\mathcal{B}\Big({m_{\chi_{k}^0}^2
	\over m_{\tilde{U}_i}^2}\Big),
\nonumber\\&&
d_{\chi_{k}^0}^g(u^I)={g_3\alpha_{s}\over32\pi e^2}\sum\limits_{i=1}^6\sum\limits_{k=1}^8{\bf Im}\Big((A_{L}^U)_{k,i}(A_R^U)_{i,k}^\dagger\Big)
{m_{\chi_{k}^ 0}\over m_{\tilde{U}_i}^2}\mathcal{B}\Big({m_{\chi_{k}^0}^2
	\over m_{\tilde{U}_i}^2}\Big).
\label{dx0u}
\end{eqnarray}
According to Eq.(\ref{LX01}) and Eq.(\ref{LX02}), $A_L^U$ and $A_R^U$ can be expressed as
\begin{eqnarray}
\nonumber&&(A_L^U)_{k,i}=-\frac{\sqrt{2}}{6} g_1 (Z^*_{\tilde{U}})^{j,i}(Z_N)^{1,k}-\frac{\sqrt{2}}{2}g_2(Z^*_{\tilde{U}})^{j,i}(Z_N)^{2,k}-\frac{\sqrt{2}}{6} g_{YX}(Z^*_{\tilde{U}})^{j,i}(Z_N)^{5,k}\\
&&\hspace{1.8cm}- Y_{u}^j(Z^*_{\tilde{U}})^{3+j,i}(Z_N)^{4,k},
\nonumber\\&&
(A_R^U)_{k,i}=
-Y_u^j(Z^*_{\tilde{U}})^{j,i}(Z^*_N)^{4,k} +\frac{\sqrt{2}}{6}(Z^*_{\tilde{U}})^{3+j,i}\Big((3g_X+4g_{YX}) (Z^*_N)^{5,j}+4g_1(Z_N^*)^{1,k}\Big).
\end{eqnarray}

Similarly, the results for down quarks are
\begin{eqnarray}
&&
d_{\chi_{k}^0}^\gamma(d^I)=-{e\over96\pi^2} \sum\limits_{i=1}^6\sum\limits_{k=1}^8{\bf Im}\Big((A_{L}^D)_{k,i}(A_R^D)_{i,k}^\dagger\Big)
{m_{\chi_{k}^ 0}\over m_{\tilde{D}_i}^2}\mathcal{B}\Big({m_{\chi_{k}^0}^2
	\over m_{\tilde{D}_i}^2}\Big),
\nonumber\\&&
d_{\chi_{k}^0}^g(d^I)={g_3\alpha_{s}\over32\pi e^2}\sum\limits_{i=1}^6\sum\limits_{k=1}^8{\bf Im}\Big((A_{L}^D)_{k,i}(A_R^D)_{i,k}^\dagger\Big)
{m_{\chi_{k}^ 0}\over m_{\tilde{D}_i}^2}\mathcal{B}\Big({m_{\chi_{k}^0}^2
	\over m_{\tilde{D}_i}^2}\Big),
\nonumber\\\nonumber
&&(A_L^D)_{k,i}=-\frac{\sqrt{2}}{6} g_1 (Z^*_{\tilde{D}})^{j,i}(Z_N)^{1,k}+\frac{\sqrt{2}}{2}g_2(Z^*_{\tilde{D}})^{j,i}
(Z_N)^{2,k}-\frac{\sqrt{2}}{6} g_{YX}(Z^*_{\tilde{D}})^{j,i}(Z_N)^{5,k}\\
&&\hspace{1.8cm}- Y_{d}^j(Z^*_{\tilde{D}})^{3+j,i}(Z_N)^{3,k},
\nonumber\\&&
(A_R^D)_{k,i}=
-Y_d^j(Z^*_{\tilde{D}})^{j,i}(Z_N^*)^{3,k} -\frac{\sqrt{2}}{6}(Z^*_{\tilde{D}})^{3+j,i}\Big((3g_X+2g_{YX}) (Z_N^*)^{5,k}+2g_1(Z_N^*)^{1,k}\Big).
\label{dx0d}
\end{eqnarray}
The one loop function $\mathcal{B}(r)$ is defined as $\mathcal{B}(r)=[2(r-1)^2]^{-1} [1+r+2r/(r-1)\ln
r]$.

  The chargino-quark-squark couplings have the same form as those in MSSM.
The contributions from chargino-squark diagrams to quark EDMs and CEDMs are
\begin{eqnarray}
&&d_{{\chi_{k}^\pm}}^\gamma(q^I)={e\over16\pi^2}
V_{{qQ}}^\dagger V_{{Qq}}
\sum\limits_{i,k}{\bf Im}\Big((A_{C}^Q)_{{k,i}}(B_{C}^Q)^\dagger
_{{i,k}}\Big){m_{{\chi_{k}^\pm}}\over m_{{\tilde{Q}_i}}^2}
\nonumber\\
&&\hspace{1.2cm}\times
\Big[-\frac{1}{3}\mathcal{B}\Big({m_{{\chi_{k}^\pm}}^2\over m_{{\tilde{Q}_i}}^2}\Big)
+\mathcal{A}\Big({m_{{\chi_{k}^\pm}}^2\over
m_{{\tilde{Q}_i}}^2}\Big)\Big]
\;,\nonumber\\
&&d_{{\chi_{k}^\pm}}^g(q^I)={g_3\over16\pi^2 }
V_{{qQ}}^\dagger V_{{Qq}}
\sum\limits_{i,k}{\bf Im}\Big((A_{C}^Q)_{{k,i}}(B_{C}^Q)^\dagger
_{{i,k}}\Big)
{m_{{\chi_{k}^\pm}}\over m_{{\tilde{Q}_i}}^2}
\mathcal{B}\Big({m_{{\chi_{k}^\pm}}^2\over m_{{\tilde{Q}_i}}^2}\Big)\;.
\label{dxzf1}
\end{eqnarray}
Here $V$ is the CKM matrix, $m_{{\chi_{k}^\pm}}~(k=1,~2)$ denote the chargino masses, Q can represent down type or up type squarks, q can represent down type or up type quarks. For down type squark diagrams,
\begin{eqnarray}\nonumber
&&(A_{C}^D)_{{k,i}}=
Y_u^j(Z_{\tilde{D}})^{j,i} (Z_{+})^{2,k},\\
&&
(B_{C}^D)_{{k,i}}=
Y_d^j(Z_{\tilde{D}})^{j+3,i} (Z_{-})^{2,k}-(Z_{\tilde{D}})^{j,i} (Z_{-})^{1,k}.
\end{eqnarray}
For up type squark diagrams,
\begin{eqnarray}\nonumber
&&(A_{C}^U)_{{k,i}}=
Y_d^j(Z_{\tilde{U}})^{j,i} (Z_{-}^*)^{2,k},\\
&&(B_{C}^U)_{{k,i}}=
 Y_u^j(Z_{\tilde{U}})^{j+3,i} (Z_{+}^*)^{2,k}-(Z_{\tilde{U}})^{j,i} (Z_{+}^*)^{1,k},
\end{eqnarray}
where  $Z_-$ and $Z_+$ are the diagonalizing matrix for $M_{\tilde{\chi}^\pm}$, i.e. $Z_-^{\dagger}M_{\tilde{\chi}^\pm}Z_+=diag(m_{{\chi_{1}^\pm}},~m_{{\chi_{2}^\pm}})$, the concrete form of $\mathcal{A}(r)$ is
$\mathcal{A}(r)=[2(1-r)^{2}]^{-1}[3-r+2/(1-r)\ln r]$.

Finally, the gluino contribution to the quark EDM and CEDM are collected here. For up type quarks, the results are
\begin{eqnarray}
&&d_{{\tilde{g}}}^\gamma(u^I)=-{4\over9\pi}e\alpha_{s}\sum\limits
_{i=1}^6{\bf Im}\Big(( Z_{{\tilde U}})^{{I+3,i}}
(Z_{{\tilde U}}^\dagger)^{{i,I}}e^{-i\theta_{3}}\Big)
{|m_{{\tilde g}}|\over m_{{\tilde{U}_i}}^2}\mathcal{B}\Big({|m_{{\tilde g}}|^2
	\over m_{{\tilde{U}_i}}^2}\Big)
\;,\nonumber\\
&&d_{{\tilde{g}}}^g(u^I)={g_3\alpha_{s}\over4\pi}\sum\limits
_{i=1}^6{\bf Im}\Big((Z_{{\tilde U}})^{{I+3,i}} (Z_{{\tilde U}}^\dagger)^{{i,I}}e^{-i\theta_{3}}\Big)
{|m_{{\tilde g}}|\over m_{{\tilde{U}_i}}^2}\mathcal{C}\Big({|m_{{\tilde g}}|^2
	\over m_{{\tilde{U}_i}}^2}\Big)\;.
\label{dg1}
\end{eqnarray}
And the corresponding results for the down type quarks are
\begin{eqnarray}
&&d_{{\tilde{g}}}^\gamma(d^I)={2\over9\pi}e\alpha_{s}\sum\limits
_{i=1}^6{\bf Im}\Big(( Z_{{\tilde D}})^{{I+3,i}}
( Z_{{\tilde D}}^\dagger)^{{i,I}}e^{-i\theta_{3}}\Big)
{|m_{{\tilde g}}|\over m_{{\tilde{D}_i}}^2}\mathcal{B}\Big({|m_{{\tilde g}}|^2
	\over m_{{\tilde{D}_i}}^2}\Big)
\;,\nonumber\\
&&d_{{\tilde{g}}}^g(d^I)={g_3\alpha_{s}\over4\pi}\sum\limits
_{i=1}^6{\bf Im}\Big(( Z_{{\tilde D}})^{{I+3,i}} (
	Z_{{\tilde D}}^\dagger)^{{i,I}}e^{-i\theta_{3}}\Big)
{|m_{{\tilde g}}|\over m_{{\tilde{D}_i}}^2}\mathcal{C}\Big({|m_{{\tilde g}}|^2
	\over m_{{\tilde{D}_i}}^2}\Big)\;,
\label{dg2}
\end{eqnarray}
 $\theta_{3}$ denotes the phase of
the gluino mass $m_{{\tilde g}}$, and the loop function $\mathcal{C}(r)=[6(r-1)^2]^{-1}[10r-26-(2r-18)/(r-1)\ln r]$.

The gluonic Weinberg operator $O_{5}$ is calculated from the two-loop "gluino-squark" diagram\cite{EDMC5}. Neglecting the contribution from b quark, the Wilson coefficient $C_{5}$ reads as
\begin{eqnarray}
&&C_{5}=-3\alpha_{s}m_{t}\Big({g_3\over4\pi}\Big)^3
{\bf Im}\Big((Z_{{\tilde T}})^{{2,2}}(Z_{\tilde {T}}^{\dagger})^{2,1}\Big)
{m_{{\tilde{T}_1}}^2-m_{{\tilde{T}_2}}^2\over |m_{{\tilde g}}|^5}
H({m_{{\tilde{T}_1}}^2\over |m_{{\tilde g}}|^2},{m_{{\tilde{T}_2}}^2
\over |m_{{\tilde g}}|^2},{m_{t}^2\over |m_{{\tilde g}}|^2}).
\label{c5}
\end{eqnarray}
Here, the $H$ function can be found in Refs.\cite{EDM2,EDM3}. $ Z_{{\tilde T}}$ is the matrix to diagonalize the mass squared matrix of stop.

The results obtained at the matching scale $\Lambda$ should be transformed down to
the chirality breaking scale $\Lambda_{\chi}\simeq1.19$ GeV \cite{EDM6,EDM7}. So the RGEs for the Wilson coefficients of the Weinberg operator and the quark EDMs, CEDMs should be solved.
In this work, the three CPV operators are considered, and they contribute to the neutron EDM:
1. the quark electric dipole operator $\mathcal{O}_\gamma=\frac{1}{4}\bar{q}\sigma_{\mu\nu}q\tilde{F}^{\mu\nu}$,
2. the quark color dipole operator $\mathcal{O}_q=\frac{1}{4}\bar{q}\sigma_{\mu\nu}T_aq\tilde{G}_a^{\mu\nu}$,
3. Weinberg's gluonic operator $\mathcal{O}_5$.
The CPV Lagrangian  $\mathcal{L}_{CPV}=\sum_i C_i(\mu)\mathcal{O}_i(\mu)$ should not depend on the scale $\mu$.
After caculation, the following relations for the coefficients are obtained\cite{EDM6}
\begin{eqnarray}
&&C_5(\mu)=\mathcal{K}^{\gamma_{GG}/\varrho} C_5(M),~~~
C_\gamma(\mu)=\mathcal{K}^{\gamma_{q}/\varrho} C_\gamma(M),\nonumber\\&&
C_q(\mu)=\mathcal{K}^{\gamma_{qq}/\varrho}C_q(M)
+C_{5}(M)\frac{\gamma_{Gq}m_q(M)}{\gamma_{qq}+\gamma_m-\gamma_{GG}}(\mathcal{K}^{\gamma_{qq}/\varrho}-\mathcal{K}^{(\gamma_{qq}-\gamma_m)/\varrho}),
\end{eqnarray}
with the parameters\cite{EDM6}
\begin{eqnarray}
&& \mathcal{K}=\frac{g_s(\mu)}{g_s(M)}, ~~~\gamma(\mathcal{O}_q)=\gamma_{qq}=\frac{29-2N_f}{3},~~~\gamma_q=\frac{8}{3},
\nonumber\\&&\gamma_{GG}=-3-2N_f, ~~~\gamma_{Gq}=6, ~~~\gamma_m=-8,~~~ \varrho=\frac{33-2N_f}{3}.\label{chz4}
\end{eqnarray}
Here, $N_f$ is the number of light quarks at scale $\mu$.

Using the values of $C_\gamma(\Lambda),~ C_q(\Lambda)$ and $ C_5(\Lambda)$, physicists obtain the corresponding contributions to quark EDM $d_q$.
At a low scale  $\Lambda_{\chi}$, the quark EDM can be obtained from $d_{q}^\gamma, d_{q}^g$ and $C_{5}(\Lambda_{\chi})$ by
the following formula\cite{EDM8}
\begin{eqnarray}
&&d_{q}=d_{q}^\gamma+{e\over4\pi}d_{q}^g+{e\Lambda_{\chi}
	\over4\pi}C_{5}(\Lambda_{\chi})\;.
\label{dq}
\end{eqnarray}
with
\begin{eqnarray}
&&d_{q}^\gamma=C_\gamma(\Lambda_{\chi})=\eta^{ED}C_\gamma(\Lambda),~~~~
d_{q}^g=C_q(\Lambda_{\chi})=\eta^{CD}C_q(\Lambda),\nonumber\\&&
C_{5}(\Lambda_{\chi})=\eta^GC_5(\Lambda).
\end{eqnarray}

From the above formulas, one can obtain the numerical results
\begin{eqnarray}
\eta^{ED}=1.53, ~~~\eta^{CD}=3.4,~~~ \eta^G=3.4. \label{shu}
\end{eqnarray}

As discussed in literature\cite{EDM6}, these numerical results of $\eta^{ED}, ~\eta^{CD}$ and $\eta^G$
in Eq.(\ref{shu}) are applicative in MSSM. 
As the U(1) extension of MSSM, B-LSSM includes two Higgs singlets and
three-generation right-handed neutrinos, where the parameters in Eq.(\ref{shu}) are also used to study neutron EDM\cite{EDM7p}.
Comparing with MSSM, $U(1)_X$SSM has additional fields: three Higgs singlets and
three-generation right-handed neutrinos.  The conditions for $\eta^{ED}, ~\eta^{CD}$ and $\eta^G$ in $U(1)_X$SSM
are same as those in B-LSSM\cite{EDM7p}. Furthermore, for the parameters in the Eq.(\ref{chz4}),
$U(1)_X$SSM does not have obvious deference from the condition of MSSM.

From the quark model, the EDM of neutron is derived from u quark's EDM $d_{u}$ and d
quark's EDM $d_{d}$ with the following expression
\begin{eqnarray}
&&d_{n}={1\over3}(4d_{d}-d_{u}).
\label{dn}
\end{eqnarray}

\section{numerical results}
In this section, we calculate the numerical results.
Manifold low energy experiments and LHC all give constraints to the $Z'$ boson properties in $U(1)_X$SSM\cite{NR0,NR0plus}. To satisfy the constraint of $Z'$ boson mass, we take $M_{Z^\prime}>5.1 ~{\rm TeV}$. According to Refs.\cite{NR1,NR2}, the constraint for the ratio between $M_{Z^\prime}$ and its gauge coupling $g_X$ is $M_{Z^\prime}/g_{X}\geq 6 ~{\rm TeV}$ at 99\% CL.
$\tan \beta_\eta$ should be smaller than 1.5, which is obtained from the LHC experimental data\cite{NR3,NR4}.
According to the research in literature\cite{gluino}, we take the mass of gluino more than 2 TeV.
Moreover, the lightest CP-even Higgs mass is a very strict experimental constraint and the latest experimental data is $m_h=125.10\pm0.14$ GeV \cite{PDG}.
We take into account all these constraints and adopt the following parameters
\begin{eqnarray}\nonumber
&& \tan\beta=10,~ \lambda_{H}=0.25,~\kappa=1,~T_{\lambda_{C}}=-0.1~{\rm TeV},~ v_S=4.3~{\rm TeV},
\nonumber\\&&
 T_{\lambda_{H}}=0.3~{\rm TeV},~T_{U11}=T_{D11}=T_{U22}=|M_{2}|=|M_{BL}|=1~{\rm TeV},
\nonumber\\ && m^2_{\tilde{Q}{1,1}}= m^2_{\tilde{D}{11}}= m^2_{\tilde{U}{22}}
=4~{\rm TeV}^2,~B_{\mu}=B_{S}=1~{\rm TeV}^2,~ \lambda_{C}=-0.1,\nonumber\\&&|M_1|=1.2~{\rm TeV},~ l_W=4 ~{\rm TeV}^2,
~m_{\tilde{g}}=2.1~ {\rm TeV},~|M_S|=1.5~{\rm TeV}.
\label{parameter}
\end{eqnarray}
To simplify the numerical discussion, the non-diagonal elements of
$m^2_{\tilde{U}},~ m^2_{\tilde{D}},~ m^2_{\tilde{Q}},~ {T_U},~ {T_D}$ are supposed as zero.

\subsection{Neutron EDM}
From equations (\ref{dx0u}), (\ref{dx0d}), (\ref{dxzf1}), (\ref{dg1}),
(\ref{dg2}) and (\ref{c5}), we can see that the none-zero EDM can be derived from none-zero imaginary part. If the rotation matrices are taken to be real, the EDM vanishes.  The new CPV sources beyond MSSM are interesting.
In the following analysis, we focus on the CPV phases in SUSY particle mass matrices(i.e. $M_{\tilde{\chi}^0}$ Eq.(\ref{mx0}) and $M_{\tilde{\chi}^\pm}$ Eq.(\ref{mxzf})) that make the rotation matrices to be complex. We also consider other sensitive parameters that affect the EDMs and CEDMs.

In the determination of adjustable parameters, we first set typical values for the parameters (as shown in Eq. (\ref{parameter})),
and then determine the values of the newly introduced parameters in $U(1)_X$SSM
according to the CPV phase angles in $M_{\tilde{\chi}^0}$ ($M_{BL}$, $M_{BB'}$ and $M_S$).
Data analysis shows that $g_X$, $g_{YX}$, $\tan\beta'$ and $\xi$ are
sensitive to neutron and quark EDMs in the introduced parameter spaces.
Their values are strongly constrained by the upper bound of the neutron EDM  $d_n<1.6 \times10 ^ {- 26}$ e$\cdot$cm.
We fix the remaining parameters as the values in Eq. (\ref{parameter}).
We randomly scan the values of $g_X$, $g_{YX}$, $\tan\beta'$ and $\xi$ to find their value ranges fitting the experiment data of $d_n$.
The four parameters emerge in several particle mass matrixes. $g_X$ and $g_{YX}$ are gauge coupling constants. Therefore,
the four parameters ($g_X$, $g_{YX}$, $\tan\beta'$, $\xi$) can influence the theoretical value of $d_n$ obviously.

\begin{figure}[ht]
	\centering
	\begin{minipage}[b]{0.49\linewidth}
		\centering
		\includegraphics[width=85mm]{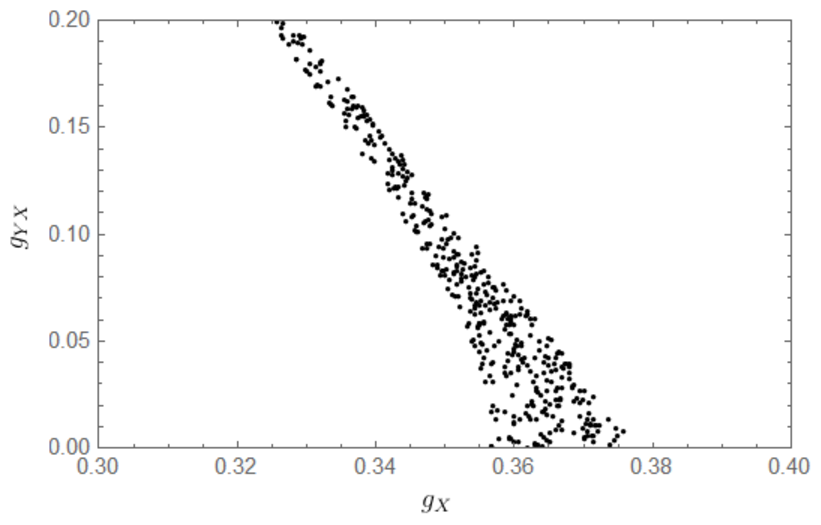}
		\caption{With $\theta_{M_{BB'}}=\pi/4,$ $\tan\beta'=0.9$ and $\xi=23$ TeV,
  the allowed region is plotted by the dots in the plane of $g_X$-$g_{YX}$.}
		\label{fig:gXgYX}
	\end{minipage}
	\quad
	\centering
	\begin{minipage}[b]{0.47\linewidth}
		\includegraphics[width=82mm]{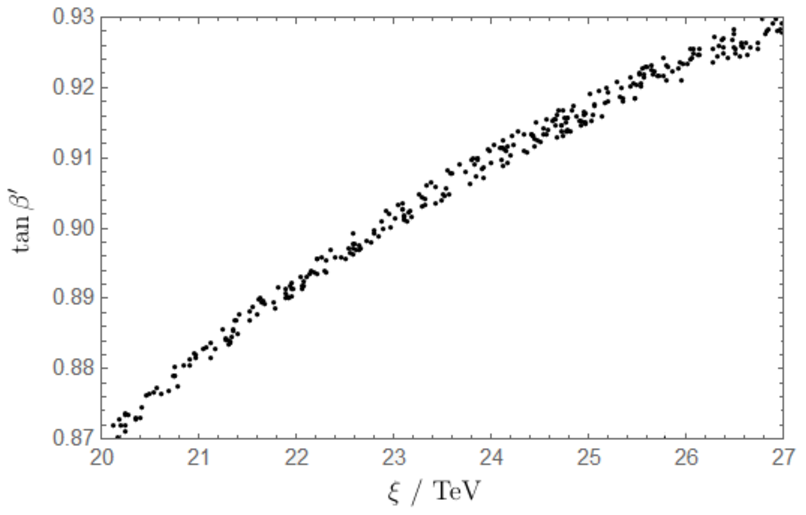}
		\caption{With $\theta_{M_{BB'}}=\pi/4,$ $g_X=0.25$ and $g_{YX}=0.1$,
  the allowed region is plotted by the dots in the plane of $\xi$-$\tan\beta'$.}
		\label{fig:VnewTanp}
	\end{minipage}
\end{figure}

With the non-zero CPV phase $\theta_{M_{BB'}}=\frac{\pi}{4}$, $\tan\beta'=0.9$ and $\xi=23$ TeV,
we plot the region in the plane of $g_X$ versus $g_{YX}$, where the dots satisfy the constraint from neutron EDM.
These results are shown in the Fig.\ref{fig:gXgYX} with $g_X$ in the region (0.3, 0.4) and $g_{YX}$ in the region (0, 0.2).
In the region of $g_X$ (0.35, 0.37) and $g_{YX}$ (0, 0.05), there are more dots. When $g_{YX}$ turns large, the allowed region shrinks.
In the Fig.\ref{fig:VnewTanp}, we show the region satisfying the limit of $d_n$ in the plane of $\tan\beta'$ and $\xi$. To obtain these results,
we use the parameters $g_X=0.25$, $g_{YX}=0.1$ and $\theta_{M_{BB'}}=\frac{\pi}{4}$. The allowed region seems a narrow band with $\xi$ varying
from 20 TeV to 27 TeV. It is well known that the smaller CPV phase angle leads to larger reasonable parameter space.

\begin{figure}[ht]
	\centering
	\begin{minipage}[b]{0.48\linewidth}
		\centering
		\includegraphics[width=83mm]{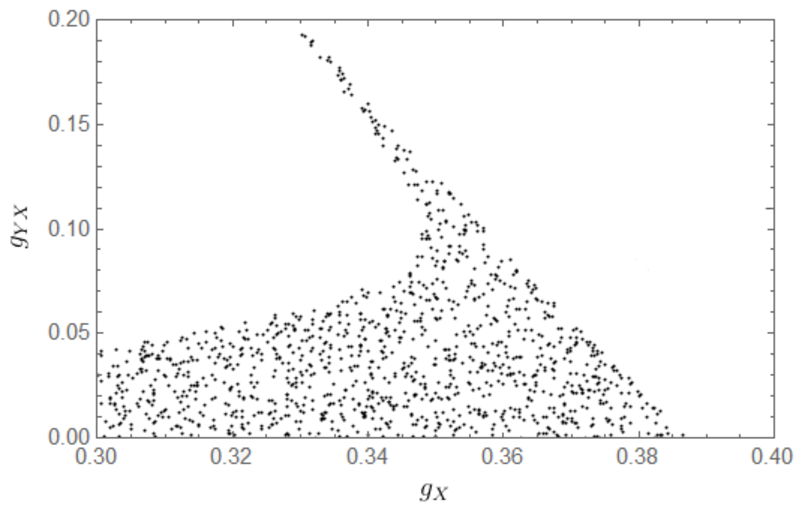}
		\caption{With $\theta_{M_{BL}}=\pi/4,$ $\tan\beta'=0.89$ and $\xi=23$ TeV,
  the allowed region is plotted by the dots in the plane of $g_X$-$g_{YX}$.}
		\label{fig:gXgYXmbl}
	\end{minipage}
	\quad
	\centering
	\begin{minipage}[b]{0.47\linewidth}
		\includegraphics[width=85mm]{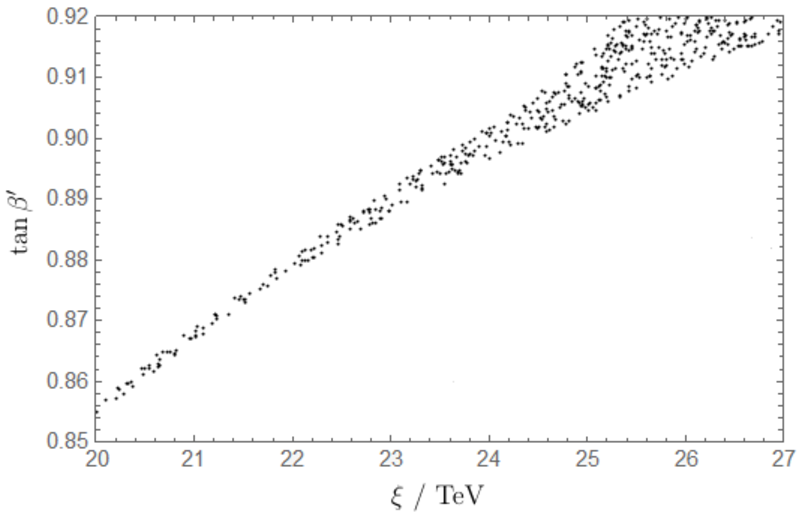}
		\caption{With $\theta_{M_{BL}}=\pi/4,$ $g_X=0.25$ and $g_{YX}=0.1$,
  the allowed region is plotted by the dots in the plane of $\xi$-$\tan\beta'$.}
		\label{fig:VnewTanpmbl}
	\end{minipage}
\end{figure}

It is interesting to study the CPV phase
beyond MSSM.
$M_{BL}$ is the new gaugino mass, which appears in the neutralino mass matrix and can
have non-zero CPV phase $\theta_{M_{BL}}$. As $\theta_{M_{BL}}=\frac{\pi}{4}$, $\tan\beta'=0.89$ and $\xi=23$ TeV,
the allowed region of $g_X$ and $g_{YX}$ is shown in the Fig.\ref{fig:gXgYXmbl}. The dots concentrate in the area with $g_X$ from 0.3 to 0.38 and
$g_{YX}$ from 0 to 0.06. As $g_{YX}>0.1$, the allowed region turns small obviously.
Similarly, the results in the plane of $\tan\beta'$ and $\xi$ are researched
with $\theta_{M_{BL}}=\frac{\pi}{4}$, $g_X=0.25$ and $g_{YX}=0.1$.
The reasonable results denoted by dots are given out in the Fig.\ref{fig:VnewTanpmbl}. Larger reasonable region is
around $\tan\beta'$ $(0.9\sim0.92)$ and $\xi$ $(24\sim 27)$ TeV.
The effects from $\theta_{M_{S}}$ are much smaller than
the effects from $\theta_{M_{BL}}$ and $\theta_{M_{BB'}}$.
That is to say, the reasonable parameter spaces
obtained from $\theta_{M_{BB'}}\neq0$ and $\theta_{M_{BL}}\neq0$ will automatically fit the condition with $\theta_{M_{S}}\neq0$.
Therefore, we do not analyze $\theta_{M_{S}}$ particularly.

\begin{figure}[ht]
 	\centering
 	\includegraphics[width=100mm]{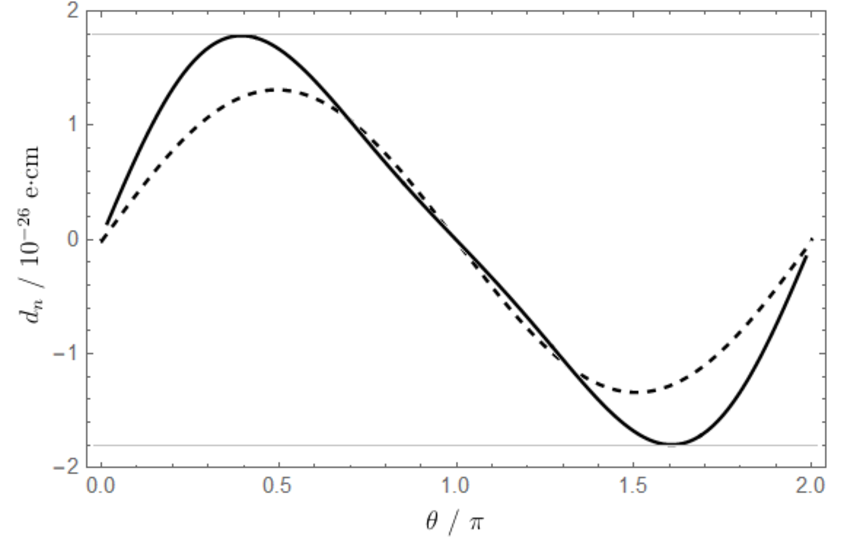}
 	\caption{$d_n$ versus $\theta_{M_{BB'}}$ with $\tan\beta'=0.9$ represented by the solid line  and $d_n$ versus $\theta_{M_{BL}}$ with $\tan\beta'=0.89$ denoted by the dashed line. The gray lines indicate the experimental limitation.}
 	\label{fig:dnmblmbbp}
 \end{figure}

 To apparently see the effects of the CPV phases $\theta_{M_{BB'}}$ and $\theta_{M_{BL}}$,
we use the parameters $g_X=0.25$, $g_{YX}=0.1$ and $\xi=23$ TeV.
As $\tan\beta'=0.9$ and $\theta_{M_{BL}}=0$, the results versus $\theta_{M_{BB'}}$ in the region (0, $2\pi$) are plotted by the solid line in the Fig.\ref{fig:dnmblmbbp}.
The dashed line represents the results versus $\theta_{M_{BL}}$ with the parameters $\tan\beta'=0.89$
and $\theta_{M_{BB'}}=0$. The dashed line is a sinusoidal function while the solid line has a small deviation.
This characteristic is caused by the the CPV phase. From the mass matrix $M_{\tilde{\chi}^0}$ we can see that $\theta_{M_{BL}}$ is a common factor 
in the imaginary parts of Eq. (\ref{dx0u}) rather than $\theta_{M_{BB'}}$. As $\theta_{M_{BB'}}=0$ and $\theta_{M_{BL}}=0$,
there is no CPV effect and $d_n=0$. When $(\theta_{M_{BL}})=\frac{\pi}{2}$ and $\theta_{M_{BB'}}\simeq \frac{2\pi}{5}$,
$d_n$ reaches the biggest value.

\subsection{Heavy c,t quark EDM}

The bounds on EDMs of heavy quarks c and t are much weaker than that of neutron. The strongest constraints on
$d_c$ and $d_t$ come from the indirect limitation of $|d_c|<1.5\times10^{-21}$ e$\cdot$cm and $|d_t|<5.0\times10^{-20}$ e$\cdot$cm
given in literature \cite{edmdc,edmdt}. The numerical analysis also shows that the restriction of $d_n$
makes $d_u$ and $d_d$ in the order of $10^{-25}\sim10^{-26}$ e$\cdot$cm, which is much more strict than the upper bound of $d_c$ and
$d_t$.
As a result, in the parameter space satisfying $d_n$ limit, we have enough range to adjust the
second and third generation soft parameters which affect $|d_c|$ and $|d_t|$, although there is  difference
in the Yukawa coupling between three generations.
In this section, unless explicitly specified or
taken as a variable, the values of parameters  are consistent with the previous subsection and Eq.(\ref{parameter})
to ensure that the strong constraint from  $d_n$  is satisfied.

\begin{figure}[ht]
	\centering
	\begin{minipage}[b]{0.99\linewidth}
		\includegraphics[width=100mm]{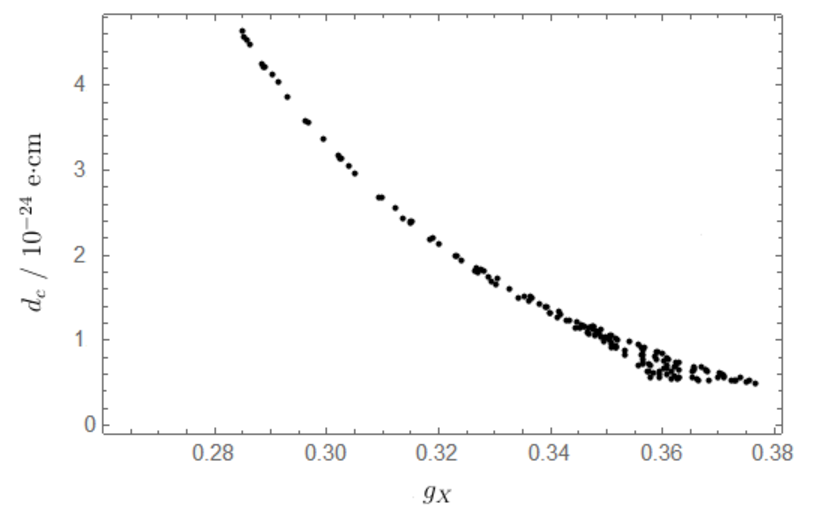}
		\caption{ Adopting $\theta_{M_{BB'}}=\pi/2$ and $m^2_{\tilde{Q}22}=4 \:{\rm TeV}^2$, we plot $d_c$ versus $g_X$.
To satisfy the limit of $d_n$, $g_{YX}$ changes accordingly with the varying $g_X$. Therefore, a series of dots are obtained.}
		\label{fig:dcgx}
	\end{minipage}
\end{figure}

 $g_X$ is the new gauge coupling constant which is sensitive to $d_c$ and $d_t$.
Its influences on $d_c$ and $d_t$ are complex because it can affect both the scalar quarks and neutralinos.
With the only non-zero CPV phase $\theta_{M_{BB'}}=\pi/2$, we study $d_c$ versus $g_X$. As discussed in the previous subsection,
the limit on $d_n$ is very strict and $g_X$ produces obvious effect to neutron EDM. Then, when we change the value of $g_X$,
the value of $g_{YX}$ should change accordingly. Based on the data in Fig.\ref{fig:gXgYX}, we plot the numerical results by the dots in the Fig.\ref{fig:dcgx}. It is obvious that $d_c$ is the decreasing function of
 $g_X$. In this figure, the region of $d_c$ is $(0.5, 4.8)\times 10^{-24}$ e$\cdot$cm. In the whole,  $\theta_{M_{BB'}}$ contribution to
 $d_c$ is at the order of  $10^{-24}$ e$\cdot$cm.

\begin{figure}[ht]
	\centering
	\begin{minipage}[b]{0.99\linewidth}
		\includegraphics[width=100mm]{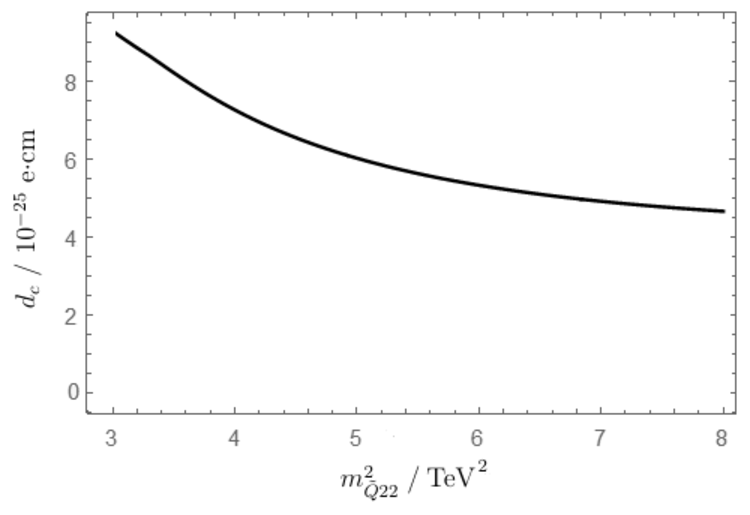}
		\caption{ Adopting $\theta_{M_{BB'}}=\pi/2$, $g_X=0.35$
and $g_{YX}=0.1$, we plot $d_c$ versus $m^2_{\tilde{Q}22}$ by the solid line.}
		\label{fig:dcmq}
	\end{minipage}
\end{figure}

$d_c$ should be influenced appreciably by the scalar charm quark masses
which have the important parameter $m^2_{\tilde{Q}22}$.  With the parameters $\theta_{M_{BB'}}=\pi/2$, $g_X=0.35$ and $g_{YX}=0.1$,
we plot $d_c$ versus $m^2_{\tilde{Q}22}$ by the solid line in the Fig.\ref{fig:dcmq}. As $m^2_{\tilde{Q}22}$
changes from $3 ~{\rm TeV}^2$ to $8~ {\rm TeV}^2$, $d_c$ turns small apparently. This characteristic implies that
our results satisfy the decoupling rule. The increase of heavy scalar quark mass should suppress the SUSY contribution to heavy quark EDM.

\begin{figure}[ht]
	\centering
	\begin{minipage}[b]{0.99\linewidth}
		\includegraphics[width=100mm]{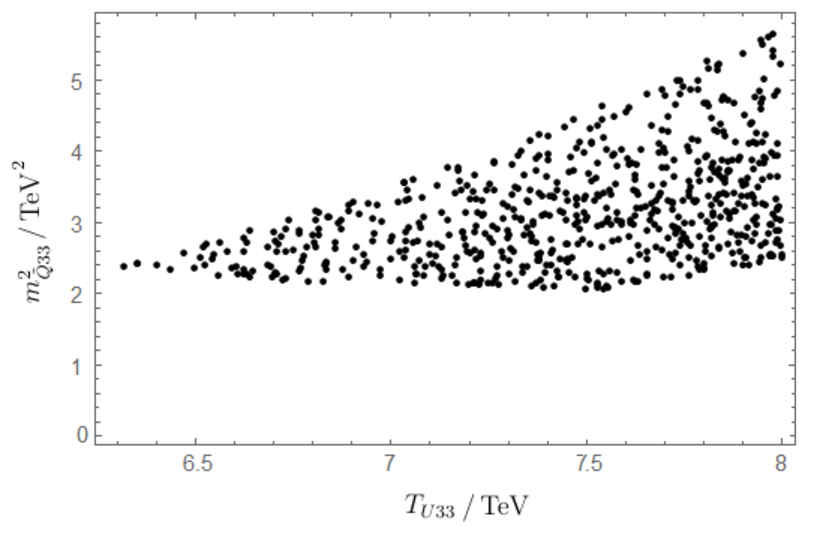}
		\caption{In the plane of $T_{U33}$ and $m^2_{\tilde{Q}{33}}$, the numerical results satisfying $123{\rm GeV}<m_h<127 {\rm GeV}$ are plotted by the dots.}
		\label{fig:125}
	\end{minipage}
\end{figure}

The third generation parameters $m^2_{\tilde{Q}{33}}$, $m^2_{\tilde{U}{33}}$ and $T_{U33}$ are sensitive to $d_t$.
 On the other hand, the t quark and scalar t quark give a major one loop contribution to the lightest CP-even Higgs mass,
and thus the bound on the lightest CP-even Higgs mass must be taken into account.
To explore the parameter space well, we random scan three parameters in the regions
$1.5 ~{\rm TeV}^2<m^2_{\tilde{Q}{33}}<7~ {\rm TeV}^2$, $1.5 ~{\rm TeV}^2<m^2_{\tilde{U}{33}}<7 ~{\rm TeV}^2$ and
$1 ~{\rm TeV}<T_{U33}<8~ {\rm TeV}$. If the numerical results for the lightest CP-even Higgs mass $m_h$ satisfy
$123~{\rm GeV}<m_h<127~{\rm GeV}$, we plot the allowed dots in the plane of $T_{U33}$ and $m^2_{\tilde{Q}{33}}$,
which is shown in the Fig.\ref{fig:125}. Because $m_h $ in three $\sigma$ is very strict, we use a little loose
restriction. The allowed region seems as a triangle with $T_{U33}$ varying from 6.3 TeV to 8 TeV and
$m^2_{\tilde{Q}{33}}$ varying from 2 ${\rm TeV}^2$ to 6 ${\rm TeV}^2$.

\begin{figure}[ht]
	\centering
	\begin{minipage}[b]{0.99\linewidth}
		\includegraphics[width=100mm]{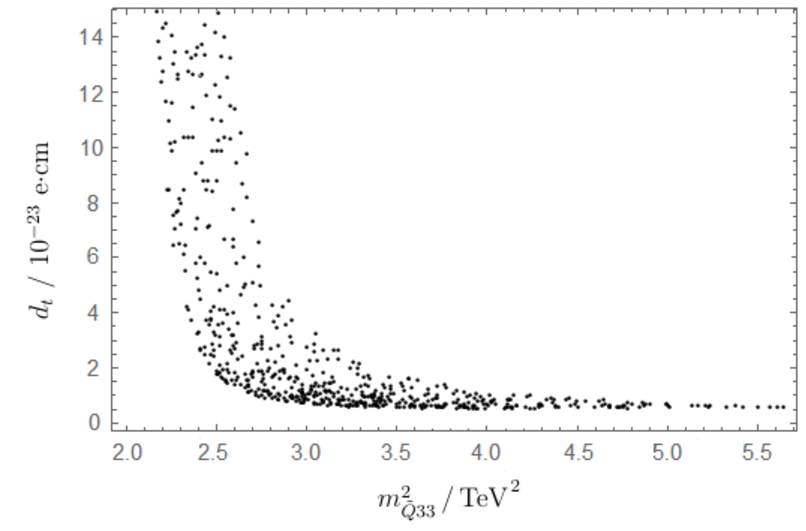}
		\caption{As $\theta_{M_{BB'}}=\pi/2$, we plot $d_t$ versus $m^2_{\tilde{Q}33}$ with
$T_{U33}$ and $m^2_{\tilde{U}33}$ changing accordingly to satisfy the constraint from $m_h$.}
		\label{fig:dtMQ33}
	\end{minipage}
\end{figure}

 The values of $T_{U33}$, $m^2_{\tilde{Q}33}$ and $m^2_{\tilde{U}33}$ can not change independently
 due to the constraint from the lightest CP-even Higgs mass.
 With the parameters satisfying the condition $123{\rm GeV}<m_h<127 {\rm GeV}$, we plot $d_t$
 versus $m^2_{\tilde{Q}33}$ by the dots in the Fig.\ref{fig:dtMQ33}.
 When $m^2_{\tilde{Q}33}$ turns from 2.2 ${\rm TeV}^2$ to 5.7 ${\rm TeV}^2$,
 the results of $d_t$ vary from $15\times10^{-23}$ e$\cdot$cm to $1\times10^{-23}$ e$\cdot$cm.
In $m^2_{\tilde{Q}33}$ region (2.2, 3.5) ${\rm TeV}^2$, $d_t$ decreases quickly.
The decreasing trend of $d_t$ turns gently, when $m^2_{\tilde{Q}33}$ is larger
than 3.5 ${\rm TeV}^2$. This diagram shows the decoupling rule evidently with the enlarging $m^2_{\tilde{Q}33}$.
Comparing with $d_c$ in the Fig.\ref{fig:dcgx}, $d_t$ in the Fig.\ref{fig:dtMQ33}
is about $d_c$ 10 to 100 times. We can declare that our numerical results satisfy the decoupling rule  from the Fig.\ref{fig:dcgx} and Fig.\ref{fig:dtMQ33}.

\section{Summary}

The $U(1)_X$SSM is obtained from extending the MSSM with the local gauge group $U(1)_X$ and the new superfields i.e. three right-handed neutrinos and three Higgs singlets  $\hat{\eta},~\hat{\bar{\eta}},~\hat{S}$.
As a extended SUSY model, the $U(1)_X$SSM has more CPV sources than the MSSM. 
The new introduced CPV phase $\theta_{M_{BB'}}$ and $\theta_{M_{BL}}$ which contribute to the EDMs of neutron and heavy quarks c, t are studied.

For convenience of calculation, we use the effective theory to describe loop-induced contributions.
The one loop contributions from the possible virtual particle diagrams to the quark EDMs and CEDMs are derived.
In the numerical results, considering the constraints from the experimental data, we study the CPV phase angles $\theta_{M_{BB'}}, ~\theta_{M_{BL}}$ with non-zero values.
Although the constraint on $d_n$ is very strict, the theoretical prediction value can be controlled well below the experimental limits with reasonable CPV phase angles and other parameters.
The numerical results for $d_c$ can reach $5.0\times10^{-24}$ e$\cdot$cm, which is about two orders smaller than its indirect constraint.
In the calculation of $d_t$, we consider the constraint of 125 GeV Higgs mass.
The largest value of $d_t$ is enhanced to $10^{-22}$ e$\cdot$cm, and this large value is about one to two order larger than $d_c$ and one order less than its indirect limit. In the near future, the updated EDMs of heavy quarks and neutron may be detected and uncover the nature of CPV.

\begin{acknowledgments}
	This work is supported by National Natural Science Foundation of China (NNSFC)
	(No. 11535002, No. 11705045), Natural Science Foundation of Hebei Province
	(A2020201002).
\end{acknowledgments}

\appendix
\section{Used mass matrices in $U(1)_X$SSM}
In the basis $ \left(\tilde{W}^-, \tilde{H}_d^-\right), \left(\tilde{W}^+, \tilde{H}_u^+\right)$, the definition of the mass matrix for charginos is given by
\begin{equation}
M_{\tilde{\chi}^\pm} = \left(
\begin{array}{cc}
M_2 &\frac{1}{\sqrt{2}} g_2 v_u \\
\frac{1}{\sqrt{2}} g_2 v_d  &\frac{1}{\sqrt{2}} {\lambda}_{H} v_S  + \mu\end{array}
\right).
\label{mxzf}
\end{equation}

In the basis $\left(\tilde{d}^0_{L,{{\alpha_1}}}, \tilde{d}^0_{R,{{\alpha_2}}}\right), \left(\tilde{d}^{0,*}_{L,{{\beta_1}}}, \tilde{d}^{0,*}_{R,{{\beta_2}}}\right)$, the definition of the squared mass matrix for down type squark is given by
\begin{equation}
M^2_{\tilde{D}} = \left(
\begin{array}{cc}
m_{\tilde{d}_L^0\tilde{d}_L^{0,*}} &m^\dagger_{\tilde{d}_R^0\tilde{d}_L^{0,*}}\\
m_{\tilde{d}_L^0\tilde{d}_R^{0,*}} &m_{\tilde{d}_R^0\tilde{d}_R^{0,*}}\end{array}
\right),
\end{equation}
where
\begin{eqnarray}
&&m_{\tilde{d}_L^0\tilde{d}_L^{0,*}} = \frac{1}{24}
 \Big( (3 g_{2}^{2}  + g_{1}^{2} + g_{Y X}^{2}) ( v_{u}^{2}- v_{d}^{2}  ) +  g_{Y X} g_{X}  (2 v_{\bar{\eta}}^{2}  -2 v_{\eta}^{2}  - v_{d}^{2}  + v_{u}^{2})\Big)+m_{\tilde{Q}}^2  +\frac{ v_{d}^{2}}{2} {Y_{d}^{\dagger}  Y_d},\nonumber\\
&&m_{\tilde{d}_L^0\tilde{d}_R^{0,*}} = -\frac{1}{2}  \Big(\sqrt{2}  (- v_d T_d  + v_u Y_d \mu^* ) + v_u v_S Y_d {\lambda}_{H}^* \Big),\nonumber\\
&&m_{\tilde{d}_R^0\tilde{d}_R^{0,*}} = \frac{1}{24}   \Big(2  (g_{1}^{2} + g_{Y X}^{2}) ( v_{u}^{2}  - v_{d}^{2})+ g_{Y X} g_{X} (4 v_{\bar{\eta}}^{2}  - 4 v_{\eta}^{2}  +5 v_{u}^{2}  - 5 v_{d}^{2} ),\nonumber \\
&&\hspace{1.8cm}+3 g_{X}^{2} (2 v_{\bar{\eta}}^{2}  - 2 v_{\eta}^{2}  + v_{u}^{2}  - v_{d}^{2})\Big) +m_{\tilde{D}}^2  +\frac{ v_{d}^{2}}{2} {Y_{d}^{\dagger}  Y_d}.\nonumber
\end{eqnarray}

In the basis $\left(\tilde{u}^0_{L,{{\alpha_1}}}, \tilde{u}^0_{R,{{\alpha_2}}}\right), \left(\tilde{u}^{0,*}_{L,{{\beta_1}}}, \tilde{u}^{0,*}_{R,{{\beta_2}}}\right)$, the definition of the squared mass matrix for up type squark is
\begin{equation}
M^2_{\tilde{U}} = \left(
\begin{array}{cc}
m_{\tilde{u}_L^0\tilde{u}_L^{0,*}} &m^\dagger_{\tilde{u}_R^0\tilde{u}_L^{0,*}}\\
m_{\tilde{u}_L^0\tilde{u}_R^{0,*}} &m_{\tilde{u}_R^0\tilde{u}_R^{0,*}}\end{array}
\right),
\end{equation}
where
\begin{eqnarray}
	&&m_{\tilde{u}_L^0\tilde{u}_L^{0,*}} = \frac{1}{24}  \Big( (g_{1}^{2} -3 g_{2}^{2}+ g_{Y X}^{2}) ( v_{u}^{2}- v_{d}^{2}  ) +   g_{Y X} g_{X}  (2 v_{\bar{\eta}}^{2}  -2 v_{\eta}^{2}  - v_{d}^{2}  + v_{u}^{2})\Big)
+  m_{\tilde{Q}}^2  +\frac{ v_{u}^{2}}{2} {Y_{u}^{\dagger}  Y_u},\nonumber\\
	&&m_{\tilde{u}_L^0\tilde{u}_R^{0,*}} = -\frac{1}{2}   \Big(\sqrt{2}  (v_d Y_u \mu^*  - v_u T_u ) + v_d v_S Y_u {\lambda}_{H}^* \Big),\nonumber\\
	&&m_{\tilde{u}_R^0\tilde{u}_R^{0,*}} = \frac{1}{24}   \Big(4  (g_{1}^{2} + g_{Y X}^{2}) (- v_{u}^{2}  + v_{d}^{2})+  g_{Y X} g_{X}  (7 v_{d}^{2}  -7 v_{u}^{2}  -8 v_{\bar{\eta}}^{2}  + 8 v_{\eta}^{2} )\nonumber \\
	&&\hspace{1.8cm}+3  g_{X}^{2} (-2 v_{\bar{\eta}}^{2}  + 2 v_{\eta}^{2}  - v_{u}^{2}  + v_{d}^{2})\Big) +  m_{\tilde{U}}^2  +\frac{ v_{u}^{2}}{2} {Y_{u}^{\dagger}  Y_u}.\nonumber
\end{eqnarray}

\end{document}